\documentclass{elsart}

 \usepackage{graphicx}
\usepackage{amssymb}

\journal{Physica A}

\begin{document}

\begin{frontmatter}

% Title, authors and addresses

% use the thanksref command within \title, \author or \address for footnotes;
% use the corauthref command within \author for corresponding author footnotes;
% use the ead command for the email address,
% and the form \ead[url] for the home page:
% \title{Title\thanksref{label1}}
% \thanks[label1]{}
% \author{Name\corauthref{cor1}\thanksref{label2}}
% \ead{email address}
% \ead[url]{home page}
% \thanks[label2]{}
% \corauth[cor1]{}
% \address{Address\thanksref{label3}}
% \thanks[label3]{}

\title{The random field Ising model with an asymmetric and anisotropic trimodal probability distribution}

% use optional labels to link authors explicitly to addresses:
% \author[label1,label2]{}
% \address[label1]{}
% \address[label2]{}

\author{Ioannis A. Hadjiagapiou\corauthref{cor1}}
\corauth[cor1]{Corresponding author.}
\ead{ihatziag@phys.uoa.gr}

\address{Section of Solid State Physics, Department of Physics,
University of Athens, Panepistimiopolis, GR 15784 Zografos,
Athens, Greece}

\begin{abstract}

The Ising model in the presence of a random field, drawn from the
asymmetric and anisotropic trimodal probability distribution
$P(h_{i})=p\; \delta(h_{i}-h_{0}) + q \delta (h_{i}+ \lambda
*h_{0}) + r \delta (h_{i})$, is investigated. The partial
probabilities $p, q, r$ take on values within the interval $[0,1]$
consistent with the constraint $p+q+r=1$, asymmetric distribution,
$h_{i}$ is the random field variable with basic absolute value
$h_{0}$ (strength); $\lambda$ is the competition parameter, which
is the ratio between the respective strength of the random
magnetic field in the two principal directions $(+z)$ and $(-z)$ and
is positive so that the random fields are competing,
anisotropic distribution. This probability distribution is an
extension of the bimodal one allowing for the existence in the
lattice of non magnetic particles or vacant sites. The current
random field Ising system displays mainly second order phase
transitions, which, for some values of $p, q$ and $h_{0}$, are
followed by first order phase transitions joined smoothly by a
tricritical point; occasionally, two tricritical points appear
implying another second order phase transition. In addition to
these points, re-entrant phenomena can be seen for appropriate
ranges of the temperature and random field for specific values
of $\lambda$, $p$ and $q$. Using the variational principle,
we write down the equilibrium equation for the magnetization
and solve it for both phase transitions and at the tricritical
point in order to determine the magnetization profile with respect
to $h_{0}$, considered as an independent variable in addition to the
temperature.

\end{abstract}
\date{\today}

\begin{keyword}
% keywords here, in the form: keyword \sep keyword
Ising model \sep asymmetric bimodal random field \sep anisotropic
interactions \sep phase-diagram \sep tricritical point \sep phase
transitions

\PACS 05.50.+q \sep 75.10.Hk \sep 75.10.Nr \sep 64.60.Kw
% PACS codes here, in the form: \PACS code \sep code
%\PACS
\end{keyword}
\end{frontmatter}

\newpage
% main text
\section{Introduction}
%\label{}

Prediction of the critical behavior of modified spin models (site
or bond diluted, random bonds, random fields), called disordered
systems, has been the subject of many studies in the last decades,
since this modification brings about considerable changes in the
critical behavior of these systems, such as replacement of a
first-order phase transition (FOPT) by a second-order phase
transition (SOPT), depression of tricritical points and critical
end points, new critical points and universality classes,
etc \cite{rsdim,rbim,huiberker}. The study of the disordered
systems is based on the standard models,
Ising, Blume-Capel, Baxter-Wu, Heisenberg, etc, modified
accordingly to meet the case under consideration. Furthermore,
extensions and versions of these models can be applied to describe
many other different situations, such as multicomponent fluids,
ternary alloys, $^{3}$He -$^{4}$He mixtures, in addition to the
magnetic systems for which these were initially conceived. The
most extensively studied model in statistical and condensed matter
physics is the spin-1/2 Ising model, since its two dimensional
version was analytically solved by Onsager (without an external
magnetic field); as a consequence, it has formed the prototype for
various generalizations. In its modified versions, it exhibits a
variety of multicritical phenomena, such as a phase diagram with
ordered ferromagnetic and disordered paramagnetic phases separated
by a transition line that changes from an SOPT to an FOPT joined
by a tricritical point (TCP); besides these, critical points,
critical end points, ordered critical points of various orders,
re-entrance can appear as in the presence of random fields. The
multicritical phenomena appear in systems presenting competition
among distinct types of ordering and there are numerous
circumstances in which this kind of phenomenon can arise. In
ferromagnetic systems in the presence of random fields, the
competition is between the parallel and random orderings, causing,
occasionally, the conversion of a continuous transition into an
FOPT and the subsequent appearance of TCP as well as re-entrance
in some cases. Random-field effects on magnetic systems have been
systematically studied not only for their own merit 
but for their experimental importance, as well.

In two dimensions an infinitesimal amount of field randomness
destroys any FOPT \cite{huiberker}. One such situation is the 
presence of random magnetic fields acting on each spin in an 
otherwise free of defects lattice; the respective pure system is 
considered to be described by Ising model, which is now transformed 
into the random field Ising model (RFIM) \cite{physicstoday,natermannvillain,imryma}.
RFIM had been the standard vehicle for studying the effects of quenched
randomness on phase diagrams and critical properties of lattice spin systems
and was studied for many years since the seminal work of Imry and
Ma \cite{imryma}. Associated with this model are the notions of
lower critical dimension, tricritical points, higher order
critical points and random field probability distribution function
(PDF). The simplest model exhibiting a tricritical phase diagram
in the absence of randomness is the Blume-Capel model -- a regular
Ising spin-$1$ model \cite{blume,capel,maletal,malakisetal}.
Although much effort has been invested in the study of the RFIM,
the only well-established conclusion is the existence of a phase
transition for $d \geq 3$ (d space dimension), that is, the
critical lower dimension $d_{l}$ is 2, resulted after a long
controversial discussion \cite{imryma,imbrie}, while many other
questions are still unanswered; among them are those of the order
of the phase transition, the existence of a tricritical point
(TCP) and the dependence of these on the form of the random field
PDF. According to the mean field approximation (MFA), the choice
of the random field PDF can lead to a continuous
ferromagnetic/paramagnetic (FM/PM) boundary as in the single
Gaussian PDF, whereas for the symmetric bimodal PDF this boundary
is divided into two parts, an SOPT branch for high temperatures
and an FOPT branch for low temperatures separated by a TCP at
$kT^{t}_{c}/(zJ)=2/3$ and $h^{t}_{c}/(zJ)=(kT^{t}_{c}/(zJ)) \times
\arg\tanh(1/\sqrt{3})\simeq 0.439$
\cite{aharony,sneiderpytte,fernandez}, where $z$ is the
coordination number, $k$ is the Boltzmann constant and $T^{t}_{c},
h^{t}_{c}$ are the tricritical temperature and random field,
respectively, so that for $T<T^{t}_{c}$ and $h>h^{t}_{c}$ the
transition to the FM phase is of first order. However, this
behaviour is not fully elucidated since in the case of the three
dimensional RFIM the high temperature series expansions yield only
continuous transitions for both PDFs \cite{gofman}; according to
Houghton et al \cite{houghton} both distributions (single Gaussian
and bimodal) predict a tricritical point with $h^{t}_{c} = 0.28
\pm 0.01$ and $T^{t}_{c} = 0.49\pm 0.03$ for the bimodal and
$\sigma^{t}_{c} = 0.36 \pm 0.01$ and $T^{t}_{c} = 0.36\pm 0.04$
for the Gaussian with critical standard deviation
$\sigma^{t}_{c}$. Galam and Birman studied the crucial issue for
the existence of a TCP within the mean field theory for a general
PDF $p(\overrightarrow{H})$ (\overrightarrow{H} random magnetic
field) by using an even degree free energy expansion up to eighth
degree in the order parameter; they proposed some inequalities
between the derivatives of the PDF up to sixth order at zero
magnetic field for the possible existence of a TCP
\cite{galambirman2}. In Monte Carlo studies for $d = 3$, Machta et
al \cite{machta}, using a Gaussian distribution, could not reach a
definite conclusion concerning the nature of the transition, since
for some realizations of randomness the magnetization histogram
was two-peaked (implying an SOPT) whereas for other ones it was
three-peaked implying an FOPT; Middleton and Fisher
\cite{middleton}, using a similar distribution for $T = 0$,
suggested an SOPT with a small order parameter exponent $\beta =
0.017(5)$; Fytas et al \cite{fytas1}, following the Wang-Landau
and Lee entropic sampling schemes for the bimodal distribution
function with random field strengths $h_{0} = 2$ and $h_{0} = 2.25$
for a simple cubic lattice found only an SOPT by
applying the Lee-Kosterlitz free energy barrier method.
Hern$\acute{a}$ndez and co-workers claim that they have found a
crossover between an SOPT and an FOPT at a finite temperature and
magnetic field for the bimodal distribution function
\cite{hernandezetal}. One of the main issues was the experimental
realization of random fields. Fishman and Aharony \cite{fishaha}
have shown that the randomly quenched exchange interaction Ising
antiferromagnet in a uniform field $H$ is equivalent to a
ferromagnet in a random field with the strength of the random
field linearly proportional to the induced magnetization. Also
another interesting result found by Galam \cite{galam3} via the
MFA was that the Ising antiferromagnets in a uniform field with
either a general random site exchange or site dilution have the
same multicritical space as the random field Ising model with the
bimodal PDF.

The usual PDF for the random field is either the symmetric bimodal

\begin{equation}
 P(h_{i})=p\delta(h_{i}-h_{0}) + q \delta (h_{i}+h_{0})  \label{bimodalp}
\end{equation}

\noindent where $p$ is the fraction of lattice sites having a
magnetic field $h_{0}$, while the rest fraction $q=1-p$ of lattice
sites has a field $(-h_{0})$ and $p = q = \frac{1}{2}$
\cite{aharony,kaufkan,andelman1}, or the Gaussian, single or
double symmetric,

\smallskip

\begin{eqnarray}
P(h_{i}) & = & \frac{1}{(2 \pi \sigma ^{2})^{1/2}} \; exp\left[-
\frac{h^{2}_{i}}{2 \sigma ^{2}}\right]             \nonumber  \\
P(h_{i}) & = & \frac{1}{2}  \frac{1}{(2 \pi \sigma^{2})^{1/2}}
\left\{exp\left[-\frac{(h_{i}-h_{0})^{2}}{2 \sigma^{2}}\right] +
exp\left[-\frac{(h_{i}+h_{0})^{2}}{2 \sigma^{2}}\right]\right\}
\label{sexp}
\end{eqnarray}

\noindent with mean value zero and ($h_{0}, -h_{0}$),
respectively, and standard deviation $\sigma$
\cite{sneiderpytte,dgaussian}.

Galam and Aharony, in a series of investigations, presented a
detailed analysis via the mean field and renormalization group of
a system consisting of $n-$component classical spins (finally
choosing $n=3$) on a $d-$dimensional lattice  of a uniaxially
anisotropic ferromagnet in a longitudinal random field extracted
from a symmetric bimodal PDF ($p=q=1/2$) without and with a
uniform magnetic field along the easy axis, respectively
\cite{galamaharony,galam1}. The uniaxial anisotropy was chosen to
be along the easy axis and the exchange couplings were of the form
$J^{(2)}=aJ^{(1)}$, where $a$ is the anisotropy and $0\leq
a\leq1$. Depending on the anisotropy (small, medium, large) a
variety of phases (longitudinal, transverse, paramagnetic),
critical points, bicritical points, and critical end points as
well as a multicritical point (an intersection of bicritical,
tricritical and critical-end-point lines) resulted. In addition to
these purely theoretical investigations, Galam proposed a model
(diluted random field) in his attempt to reproduce some of the
features in the phase diagram of the experimental sample
consisting of the mixed cyanide crystals $X(CN)_{x}Y_{1-x}$, where
$X$ stands for an alkali metal (K,Na,Rb) and $Y$ a spherical
halogen ion (Br,Cl,I); the dilution of the pure crystal $XCN$ is
achieved by replacing $CN$ by the halogen ions $Y$ \cite{galam2}.
The pure alkali-cyanide $XCN$ crystal ferroelastic transition
disappears at some concentration $x_{c}$ of the cyanide; its
numerical value depends on both components $X,Y$. By choosing a
model Hamiltonian (ferromagnetic Ising-type with nearest neighbor
interaction) with dilution and a symmetric trimodal PDF for the
random fields Galam, using MFA, managed to predict the involved
first and second order phase transitions with the interfering TCP
as well as the respective concentration for a phase transition to
occur depending on the procedure considered. The random fields
were necessary because there were experimental evidences that
below $x_{c}$ cyanide displayed orientational freezing and the
random fields were used for fixing this orientation. The involved
probability $p_{t}$ in PDF as well as the critical threshold
$x_{c}$ were expressed in terms of microscopic quantities.

Recently, the asymmetric bimodal PDF (\ref{bimodalp}) with $p\neq
q$, in general, has also been studied in detail \cite{asymmetric}
as well as the respective one with random anisotropic interactions
 $P(h_{i})=p\delta(h_{i}-h_{0}) + q \delta (h_{i}+\lambda *h_{0})$
\cite{asym2}, with the competition parameter $\lambda$ varying in
the interval $[0,1]$. The former study has revealed that for some values
of $p$ and $h_{0}$, the PM/FM boundary is exclusively of second
order; however, for some other ranges of these variables this
boundary consists of two branches, a second order one and another
of first order with an intervening TCP, thus confirming the
existence of such a point, whose temperature depends only on the
probability $p$ in (\ref{bimodalp}). In addition to these
findings, re-entrance has occurred as well as complex
magnetization profiles with respect the random field strength
$h_{0}$. For $p = q = 1/2$, the symmetric bimodal PDF, the results
found by Aharony were confirmed \cite{aharony}. In the latter
study, the anisotropic interactions (introduced through the
parameter $\lambda$, with $\lambda \in [0,1]$) do not change the
numerical value of the tricritical temperatures (they still depend
on $p$ only), whereas the TCP random field ($h^{TCP}_{0}$) as well as
auxiliary one ($V^{TCP}_{0}$) change as $\lambda$ varies. Another
important influence of $\lambda$ is to reduce the FM region allocated
to the system as  $\lambda$ tends to $1$ ($\lambda \rightarrow 1$) and
simultaneously broaden the PM region; however, the overall
structure of the phase diagram as a function of  $\lambda$ for a
specific value of $p$ is unchanged, the only influence of
$\lambda$ on it is to cause a parallel translation of the wider
phase diagram, occurring for  $\lambda=0$ inwards, towards the T
axis as $\lambda$ increases, thus reducing the FM region; the
largest reduction occurs for $\lambda =1$.

An immediate generalization of the asymmetric bimodal
(\ref{bimodalp}) is the asymmetric trimodal one,

\begin{equation}
 P(h_{i})=p\delta(h_{i}-h_{0}) + q \delta (h_{i}+h_{0}) +
 r \delta(h_{i})  \label{trimodal}
\end{equation}

\noindent where $p+q+r=1$. In earlier studies, the partial probabilities
$p, q$ had been considered as equal and related to $r$ by the
relation $p=q=(1-r)/2$, symmetric PDF \cite{trimodal,saxena};
recently unequal $p, q$, ($p \neq q$) were considered, asymmetric
PDF \cite{asymtrim}. The third-peak, introduced in addition to the
other two ones in the bimodal (\ref{bimodalp}) and associated with
the third term in (\ref{trimodal}), is to allow for the presence
of non magnetic particles or vacancies in the lattice that are not
affected by the random magnetic fields and results in reducing the
randomness of the system, as well. A direct result of the choice of
this PDF is that the respective physical system, depending on the
values of $p, q, h_{0}$, can have one tricritical point and, in
some cases, it can have two such points, in contrast to the bimodal
PDF, which has only one such point in both versions, \cite{asymmetric,asym2}.

For the critical exponents of the three-dimensional RFIM, it seems
that there is broad consensus concerning their values except for
the specific heat exponent $\alpha$, for which there is much
dispute concerning its numerical value, since its sign is widely
accepted to be negative. The main sources of information for the
critical exponents are Monte Carlo simulations. However, they
provide various values depending on the probability distribution
considered. Middleton and Fisher concluded that the
$\alpha$-exponent is near zero, $\alpha = -0.01 \pm 0.09$
\cite{middleton}. Rieger and Young, considering the bimodal
distribution, estimated $\alpha = -1.0 \pm 0.3$
\cite{riegeryoung}, Rieger, using the single Gaussian
distribution, estimated $\alpha = -0.5 \pm 0.2$ \cite{rieger},
whereas Hartmann and Young, from ground-state calculations,
estimated $\alpha = -0.63 \pm 0.07$ \cite{hartmannyoung}. Nowak et
al estimated that $\alpha = -0.5 \pm 0.2$ \cite{nowak}, whereas
Dukovski and Machta found a positive value, namely, $\alpha =
0.12$ \cite{dukovski}. Malakis and Fytas \cite{malakisfytas}, by
applying the critical minimum-energy subspace scheme in
conjunction with the Wang-Landau and broad-histogram methods for
cubic lattices, proved that the specific heat and susceptibility
are non-self-averaging using the bimodal distribution. The same
ambiguous situation prevails in experimental measurements; see
Ref. \cite{belangeryoung}.

Another possible generalization for the trimodal PDF (\ref{trimodal}) 
is to assume that the random field takes on different values in the 
up and down directions (anisotropy), namely,

\begin{equation}
 P(h_{i})=p\delta(h_{i}-h_{0}) + q \delta (h_{i}+\lambda *h_{0})
     + r \delta (h_{i})   \label{trimodalr}
\end{equation}

where $\lambda$ (competition parameter) is the ratio of
the two fields in the up and down directions with $\lambda \in
[0,1]$, since for $\lambda <0$ the two random fields will act in
the same direction without competition, see also Ref.
\cite{asym2}.

In this work, we study the RFIM with the asymmetric and
anisotropic trimodal PDF (\ref{trimodalr}) with arbitrary values
for the partial probabilities $p, q$ and $ \lambda$ in order to
investigate the phase diagrams, phase transitions, tricritical
points and magnetization profiles with respect to $h_{0}$ and
compare these results with those of the isotropic case ($\lambda
=1$) studied earlier \cite{asymtrim}. The paper is organized as
follows. In the next section, the suitable Hamiltonian is
introduced, and the respective free energy and equation of state
for the magnetization are derived. In section $3$, the phase
diagram, tricritical points and magnetization profiles for various
values of $\lambda$ and $p, q$ are calculated and discussed; we
close with the conclusions in section $4$.

\vspace{-8mm}

\section{The model}

\vspace{-5mm}

The Ising model Hamiltonian in the presence of random fields
is written as

\begin{equation}
 H=-J\sum_{<i,j>}S_{i}S_{j}-\sum_{i}h_{i}S_{i}  \hspace{2mm},
    \hspace{20mm} S_{i}=\pm1.   \label{rham}
\end{equation}

\noindent The summation in the first term extends over all nearest
neighbors and is denoted by $<i,j>$; in the second term $h_{i}$
represents the random field that couples to the one-dimensional
spin variable $S_{i}$. We also consider that $J > 0$ so that the
ground state is ferromagnetic in the absence of random fields. The
presence of randomness involves two averaging procedures,
the usual thermal average, denoted by angular brackets
$\langle...\rangle$, and the disorder average over the random
fields denoted by $\langle...\rangle_{h}$ for the respective PDF.

For the asymmetric ($p \neq q$) and anisotropic ($\lambda \neq 1$)
bimodal/trimodal PDFs, we also make additional assumptions
concerning the random field moments

\begin{equation}
 <h_{i}>_{h} = (p - \lambda \, q)h_{0},
   \hspace{20mm} <h_{i} h_{j}>_{h} = h^{2}_{0}\delta _{ij} \label{h0}
\end{equation}

\noindent The former relation in (\ref{h0}) vanishes for a
symmetric and isotropic PDF ($p=q, \lambda = 1$), whereas for an
asymmetric PDF ($p \neq q$) is non-zero implying that the system
is under the influence of a residual magnetic field due to the
asymmetry and anisotropy of the random field, thereby affecting
system's magnetization; a similar case has appeared in Ref.
\cite{asymmetric,asym2,asymtrim}, as well. The latter relation
implies that there is no correlation between $h_{i}$ at different
lattice sites.

According to the MFA the Hamiltonian (\ref{rham}) takes the form
\cite{aharony,sneiderpytte,andelman1,asymmetric,asym2,asymtrim}

\begin{equation}
H_{MFA}=\frac{1}{2} NzJM^{2} - \sum_{i}(zJM + h_{i})S_{i}
     \label{mfaham}
\end{equation}

\noindent where $N$ is the number of spins and $M$ the magnetization;
the respective free energy per spin within the MFA is

\begin{eqnarray}
\frac{1}{N}\langle F \rangle_{h} & = & \frac{1}{2} zJM^{2} -
\frac{1}{\beta} \langle \ln\{ 2 \cosh [\beta (z J M + h_{i})] \}
\rangle _{h}          \nonumber  \\
  & = &  \frac{1}{2} zJM^{2} - \frac{1}{\beta}
 \int P(h_{i})\ln\{ 2 \cosh [\beta (z J M + h_{i})] \} dh_{i}
                \label{mfafren}
\end{eqnarray}

\noindent where the probability $P(h_{i})$ is chosen to be the
modified trimodal (\ref{trimodalr}), $\beta = 1/(kT)$, $T$ is the
temperature.

The magnetization is the solution to the Eq. $d(\langle F
\rangle_{h}/N) / dM = 0$ (equilibrium condition)

\begin{equation}
  M = \langle \tanh [ \beta ( zJM + h_{i} ) ] \rangle_{h}
     \label{magnet1}
\end{equation}

If the distribution $P(h_{i})$ under consideration is symmetric,
$P(h_{i})$ = $P(-h_{i})$, which occurs for $p = q = (1-r)/2$ and
$\lambda =1$, then the case $M = 0$ (PM phase) will always be a
solution to (\ref{magnet1}); otherwise this shall not be the case.
However, this can be remedied if an auxiliary field $V_{0}$ is
introduced into the system such that
\cite{aharony,asymmetric,asym2,asymtrim}

\begin{equation}
  \langle \tanh[ \beta ( h_{i} + V_{0} ) ] \rangle_{h} = 0,
  \label{externalv}
\end{equation}

\noindent inducing the PM phase; the solution to this equation is
$V_{0}$ for specific values of $h_{i}$ and $\beta$. However, this
relation acts as a constraint on the system influencing,
nevertheless, its behaviour. The free energy (\ref{mfafren}) in
the presence of the auxiliary field $V_{0}$ takes, now, the form

\begin{eqnarray}
\frac{1}{N}\langle F \rangle_{h} & = & \frac{1}{2} zJM^{2} -
\frac{1}{\beta} \langle \ln\{ 2 \cosh [\beta (zJM + h_{i} +
V_{0})] \} \rangle_{h}          \nonumber  \\
  & = &  \frac{1}{2} zJM^{2} - \frac{1}{\beta}
 \mbox{\Large \{}\! F_{0} + \frac{\alpha ^{2} F_{2}}{2!} M^{2} +
 \frac{\alpha ^{3} F_{3}}{3!}M^{3}
 +\frac{\alpha ^{4} F_{4}}{4!} M^{4} + \nonumber  \\
&& \frac{\alpha ^{6} F_{6}}{6!}M^{6}  \mbox{\Large \}}
\label{mfafren2}
\end{eqnarray}

\noindent after expanding the quantity in angular brackets in
powers of $M$ and calculating the average values using
(\ref{trimodalr}) with $\alpha \equiv \beta Jz$. By setting $t_{i}
\equiv \tanh[\beta(V_{0}+h_{i})]$, $t_{+} \equiv
\tanh[\beta(V_{0}+h_{0})]$, $t_{-} \equiv
\tanh[\beta(V_{0}-\lambda * h_{0})]$ and $t_{0} \equiv \tanh[\beta
V_{0}]$ we get

\begin{eqnarray}
F_{0} & = & \langle \ln\{ 2\cosh[ \beta ( V_{0} + h_{i} )  ] \}
\rangle_{h}     \nonumber  \\
  & = & \ln2 + p\ln\cosh[\beta (V_{0} + h_{0})] + q\ln\cosh[\beta (V_{0} - \lambda* h_{0})]  \nonumber
      + r\ln\cosh[\beta V_{0} ]  \nonumber  \\
F_{1} & = & \langle t_{i}\rangle_{h} =  p t_{+} + q t_{-} + r t_{0}   \nonumber  \\
F_{2} & = & \langle 1 - t_{i}^{2}\rangle_{h} =  1 - p t_{+}^{2} - q t_{-}^{2}-r t_{0}^{2}  \nonumber \\
F_{3} & = & \langle -2t_{i} (1-t_{i}^{2}) \rangle_{h}   \nonumber \\
  & = & -2p t_{+} (1-t_{+}^{2}) -2 q  t_{-} (1-t_{-}^{2})-2 r t_{0}(1-t_{0}^{2})  \nonumber\\
F_{4} & = & \langle 2 (1 - t_{i}^{2}) (3t_{i}^{2}-1) \rangle_{h}   \nonumber \\
  & = & 2p (1 - t_{+}^{2}) (3t_{+}^{2}-1) + 2q (1 - t_{-}^{2}) (3t_{-}^{2}-1)   \nonumber
      + 2r (1 - t_{0}^{2}) (3t_{0}^{2}-1)   \nonumber \\
F_{6} & = & \langle 8 (1 - t_{i}^{2})(15t_{i}^{4}-15t_{i}^{2}+2)
\rangle_{h}  \nonumber \\
  & = & 8p (1 - t_{+}^{2})(15t_{+}^{4}-15t_{+}^{2}+2)+
        8q (1 - t_{-}^{2})(15t_{-}^{4}-15t_{-}^{2}+2)  \\  \nonumber
  &  &  +8r (1 - t_{0}^{2})(15t_{0}^{4}-15t_{0}^{2}+2)   \label{mfafren3}
\end{eqnarray}

The condition (\ref{externalv}), for the existence of the PM phase
for any value of $p, q, \lambda$, is equivalent to $F_{1} = 0$,

\begin{equation}
  p t_{+} + q t_{-} + r t_{0} = 0   \label{f1zero}
\end{equation}

The equilibrium magnetization is a solution to the condition
$d(\langle F \rangle_{h}/N) / dM = 0$, equivalent to

\begin{equation}
M=\alpha F_{2} M + \frac{\alpha ^{2} F_{3}}{2!} M^{2} +
   \frac{\alpha ^{3} F_{4}}{3!} M^{3} +
   \frac{\alpha ^{5} F_{6}}{5!}M^{5}            \label{eqmagn}
\end{equation}

or

\begin{eqnarray}
M & = & A M + B M^{2}+ C M^{3} + E M^{5}  \label{magnet1a} \\
A & \equiv & \alpha F_{2}, B\equiv\frac{\alpha ^{2} F_{3}}{2!},
C\equiv\frac{\alpha ^{3} F_{4}}{3!}, E\equiv\frac{\alpha ^{5}
F_{6}}{5!}         \label{magnet2}
\end{eqnarray}

In RFIM if there is a phase transition it will be associated with
the magnetization and the involved two phases are the PM with
$M=0$ and the FM with $M\neq0$. The resulting phase boundary is
found by solving Eq. (\ref{magnet1a}) in conjunction with the free
energy (\ref{mfafren2}) and condition (\ref{f1zero}). The SOPT
boundary is determined by setting $A = 1$ and $C < 0$, whereas the
FOPT boundary is determined by $A = 1$ and $C > 0$. These two
boundaries, whenever they appear sequentially for the same values
of the parameters $\lambda, p, q$, are joined at a tricritical point
determined by the condition $A = 1$ and $C = 0$
\cite{aharony,sneiderpytte,houghton,kaufkan,andelman1,asymmetric,asymtrim,khurana},
provided that $E<0$ (equivalently, $F_{6}<0$) for stability
\cite{asymmetric,asymtrim,crok2,crok3}. However, for the FOPT
boundary we shall also use the equality of the respective free
energies $F(M=0) = F(M \neq 0)$, where $F \equiv \langle F
\rangle_{h}/N$.

\section{Phase diagram. Tricritical Point. Magnetization profiles}

The TCP coordinates $(T^{TCP}, h^{TCP}_{0}, V^{TCP}_{0})$,
according to the definition of this point in the previous
paragraph, are solutions to the simultaneous equations

\begin{eqnarray}
pt_{+} + qt_{-} + rt_{0}& = & 0           \nonumber \\
pt_{+}^{2} + qt_{-}^{2} + rt_{0}^{2} + 1/\alpha & = & 1    \nonumber  \\
4(pt_{+}^{2} + qt_{-}^{2} + rt_{0}^{2})- 3(pt_{+}^{4} + qt_{-}^{4}
+ rt_{0}^{4}) & = & 1 \label{simequstrim}
\end{eqnarray}

\noindent which do not lead to analytical formulas for these
coordinates for both forms of the trimodal PDF (anisotropic and
isotropic); on the contrary, the bimodal PDF (resulting from the
trimodal by setting $r=0$), even in the presence of anisotropy,
leads to analytical formulas; namely, the respective tricritical
temperature $T^{TCP}$ satisfies the second-degree equation

\begin{equation}
 3(Q-1) \alpha ^{2}_{TCP} + 2(2-3Q)\alpha_{TCP} +3Q=0   \;\; \label{tricritequ}
\end{equation}

\noindent where $Q =(p^{3}+q^{3})/qp$, thus $T^{TCP}$ is a
function only of the probability $p$ and independent of the
competition parameter $\lambda$ \cite{asymmetric}. The relevant
discriminant of Eq. (\ref{tricritequ}) is real in the interval
$(13-\sqrt{13})/26 \cong 0.37... \leq p \leq (13+\sqrt{13})/26
\cong 0.63...$, so only for these values of $p$ there exist
tricritical points and both phase transitions take place for the
same $p$, but for different temperatures and $h_{0}$s. The two
solutions to Eq. (\ref{tricritequ}) determine the respective
tricritical temperatures (in units of ($Jz/k$)), the upper and
lower ones

\begin{equation}
\frac{kT^{TCP}_{\pm}}{Jz}  =  \frac{3(Q-1)}{3Q-2\pm(4-3Q)^{1/2}}
  \\  \label{tctemp}
\end{equation}

We retain only the minus solution $T^{TCP}_{-}$, since
the plus one $T^{TCP}_{+}$ does not lead to physical results and
is thus neglected, see Ref. \cite{asymmetric}.

The remaining coordinates $h_{0}^{TCP}$ and $V_{0}^{TCP}$ for the
bimodal PDF are

\begin{eqnarray}
h_{0}^{TCP} & = & \frac{1}{2(1+\lambda)} \ln \left[
  \left(\frac {1+z_{2}}{1-z_{2}} \right) \left(\frac {1+z_{1}}{1-z_{1}}\right)\right]
   \frac{kT_{-}^{TCP}}{zJ}   \nonumber\\
V_{0}^{TCP} & = & \frac{1}{2(1+\lambda)} \ln \left[
  \left(\frac {1+z_{2}}{1-z_{2}} \right)^{\lambda} \left(\frac {1-z_{1}}{1+z_{1}}\right)\right]
   \frac{kT_{-}^{TCP}}{zJ}     \label{h0V0}
\end{eqnarray}

where $z_{1}=\sqrt{p(\alpha^{TCP}_{-}-1)/(q\,\alpha^{TCP}_{-})}$
and $z_{2}=\sqrt{q(\alpha^{TCP}_{-}-1)/(p\,\alpha^{TCP}_{-})}$;
$h_{0}^{TCP}$ and $V_{0}^{TCP}$ depend on both parameters $p$ and
$\lambda$ (unlike $T^{TCP}_{-}$) as well as on the tricritical
temperature itself $T^{TCP}_{-}$.

In order to examine the validity of the process under
consideration, we focus on the well-studied symmetric and
isotropic bimodal PDF (\ref{bimodalp}), resulting from
(\ref{trimodalr}) by setting $p=q=1/2, r=0$ and $\lambda=1$ with
respective $Q = 1$; the plus solution vanishes ($kT^{TCP}_{+}/(Jz)
= 0$), whereas the minus solution ($kT^{TCP}_{-}/(Jz)$) is
singular; however, this singularity can be removed either by using
the de L' H\^{o}pital rules in Eq. (\ref{tctemp}) or by setting in
(\ref{tricritequ}) $Q = 1$, so that ($kT^{TCP}_{-}/(Jz)$) equals
($2/3$) in agreement with the tricritical temperature in Ref.
\cite{aharony}, see also Refs. \cite{asymmetric,asym2,asymtrim}.

\begin{figure}[htbp]
\begin{center}
\includegraphics*[height=0.75\textheight]{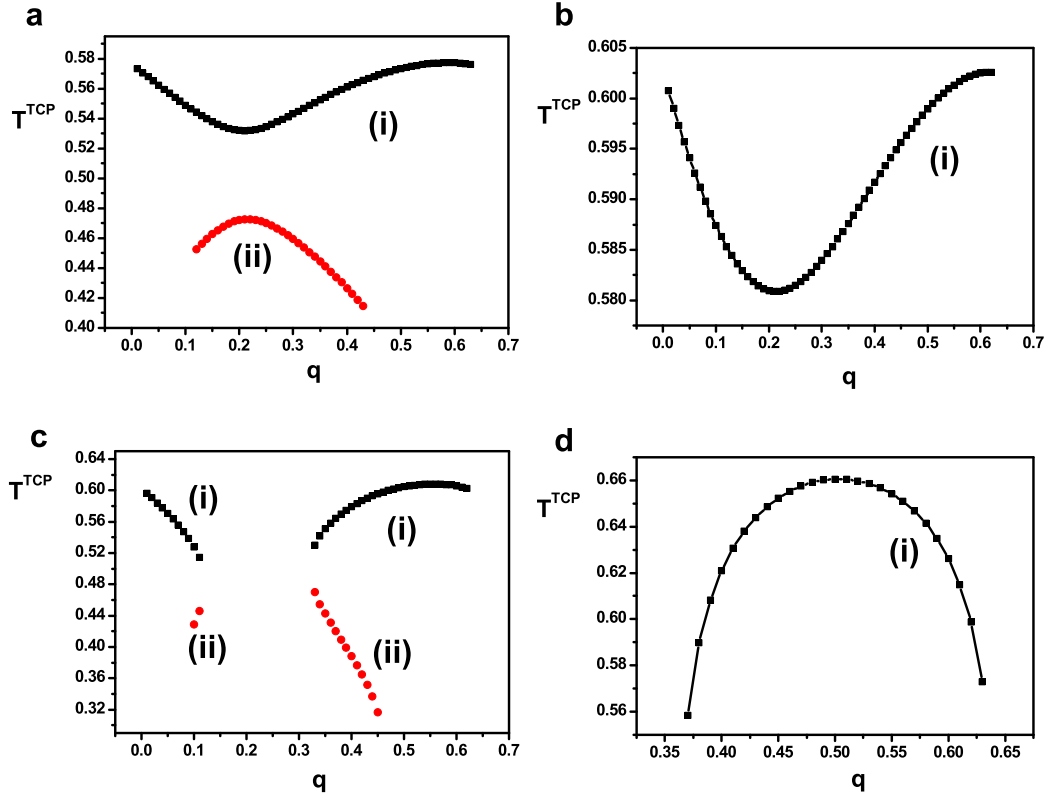}
\caption{\label{figa}(Color online)  Representative graphs for the
variation of the tricritical temperature for various values of the
competition parameter $\lambda$ and probability $p$ against the
probability $q$. The labels (i) (black symbols) and (ii) (red
symbols) refer to the upper and lower TCP temperatures,
respectively, in case two such temperatures exist for the same
$\lambda$ and $p$. Panels (a) and (b) correspond to
$\lambda=0.25$, for $p=0.37$ and $p=0.38$, respectively; panel (c)
$\lambda=0.50$ and $p=0.38$; panel (d) $\lambda=0.75$ and
$p=0.01$. $T^{TCP}$ is in units of $(Jz/k)$.}
\end{center}
\end{figure}

%\vspace{-6mm}

For the current model, the system of Eqs. (\ref{simequstrim}) can
be solved only numerically for determining the TCP coordinates;
this is achieved only for a limited number of $p$'s and $q$'s for
specific $\lambda$ values; for the TCP point, in general, $p \in
[0.0,x]$, in this interval the numerical value of the upper bound
$x$ depends upon the specific $\lambda$ value, but, in any case, it
satisfies the inequality $x \leq (13+\sqrt{13})/26 \cong 0.63...$; the
respective $q$ values depend on the specific $p$ values, but they
also lie in same interval as $p$. When $p$ takes on the value $p =
(13-\sqrt{13})/26 \cong 0.37...$ (which is the lower bound of the
probability $p$ for the bimodal PDF to have tricritical points),
then $q$ takes on all the values within the interval $[0,
(13+\sqrt{13})/26 \cong 0.63...$]; for $(13-\sqrt{13})/26 \leq p
\leq (13+\sqrt{13})/26$, the maximum possible value for $q$ is
such that $p+q = 1$ so that $r=0$. The resulting tricritical
temperatures exhibit a variety of variations as functions of the
competition parameter $\lambda$ and site probabilities $p, q$;
such graphs appear in Fig.~\ref{figa}. However, for some $p$ and
$q$ values two tricritical temperatures occur, the upper ones
(shown in black) and the lower ones (shown in red) as in
Figs.~\ref{figa}(a,c) \cite{galam,weizenmann}. In panel (c) the
TCP temperatures are grouped into two sets, the left-hand side one
for small $q$'s and the right-hand side one for larger $q$'s;
also, in this panel, the lack of points in the interval
$[0.12,0.32]$ of the $q$-axis is due to the absence of tricritical
points in this interval, thus forming the existing gap; a similar
behavior is also observed for another values of $\lambda, p, q$.

\begin{figure}[htbp]
\begin{center}
\includegraphics*[height=0.75\textheight]{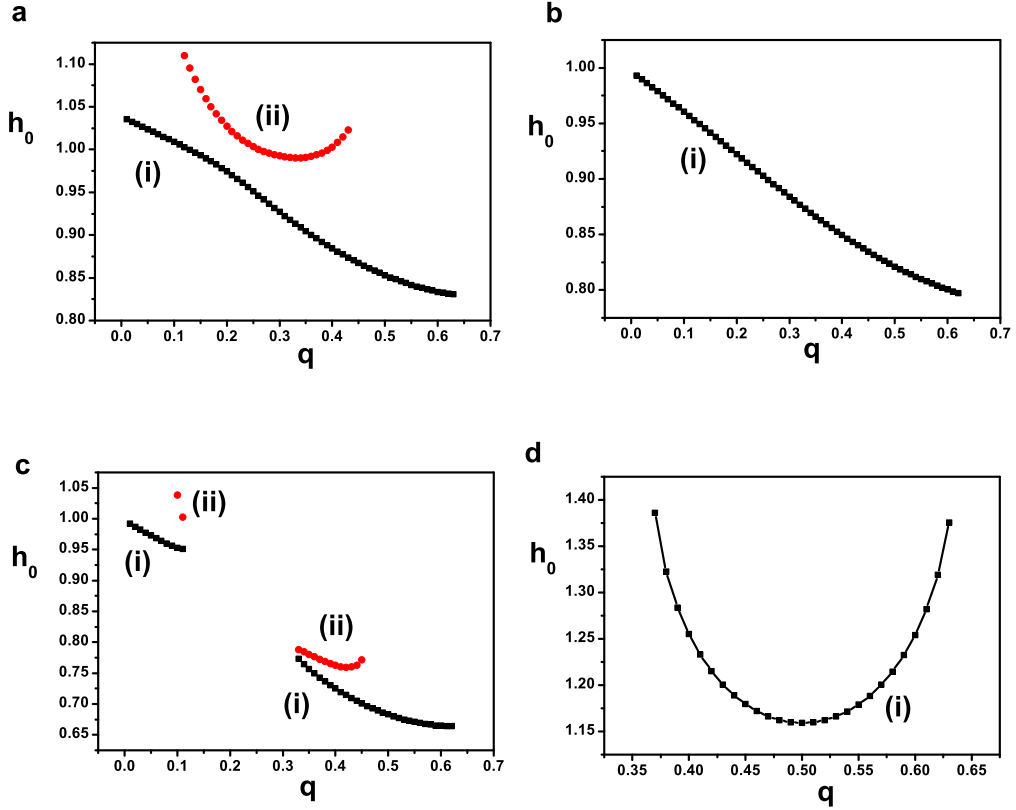}
\caption{\label{figb}(Color online) Indicative modes of variation
for the random field strength $h_{0}$ with $q$ for specific values
of $\lambda$ and $p$ at the tricritical point; the labels (i)
(black symbols) and (ii) (red symbols) refer to the quantities
corresponding to the upper and lower TCP temperatures,
respectively, in case two such temperatures exist. Panel (a)
corresponds to $\lambda=0.25, p=0.37$; panel (b) $\lambda=0.25,
p=0.38$; panel (c) $\lambda=0.50, p=0.38$; panel (d)
$\lambda=0.75, p=0.01$. The random field $h_{0}$ exhibits
monotonic and non-monotonic behavior with $q$. $h_{0}$ is in units
of $(Jz)$, i.e., $h_{0} \equiv h_{0}/(Jz)$.}
\end{center}
\end{figure}

\begin{figure}[htbp]
\begin{center}
\includegraphics*[height=0.75\textheight]{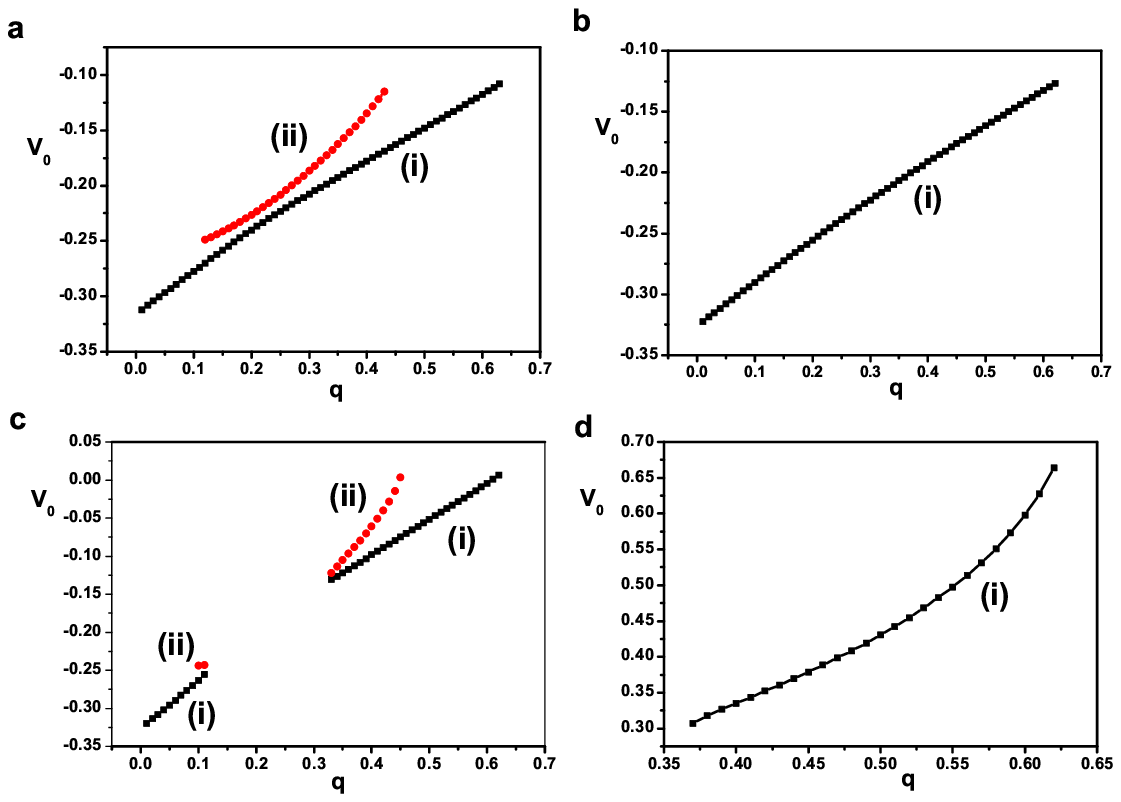}
\caption{\label{figc}(Color online) Modes of variation of the
auxiliary field $V_{0}$ with $q$ for specific values of $\lambda$
and $p$ at the tricritical point; the labels (i) (black symbols)
and (ii) (red symbols) refer to the quantities corresponding to
the upper and lower TCP temperatures, respectively, in case two
such temperatures exist. Panel (a) corresponds to $\lambda=$
$0.25, p=0.37$; panel (b) $\lambda=0.25, p=0.38$; panel (c)
$\lambda=0.50, p=0.38$; panel (d) $\lambda=0.75, p=0.01$. The
auxiliary potential $V_{0}$ exhibits monotonic behavior with $q$.
$V_{0}$ is in units of $(Jz)$, i.e., $V_{0} \equiv V_{0}/(Jz)$.}
\end{center}
\end{figure}

The variation of the random field values at the TCP,
$h_{0}^{TCP}$, resulting from Eqs. (\ref{simequstrim}) appears in
Fig.~\ref{figb}, where various modes of variation are shown,
displaying monotonic and non monotonic behavior. The gap in panel
Fig.~\ref{figb}(c) is due to the absence of tricritical points for
these values of $\lambda, p, q$ as in Fig.~\ref{figa}(c). A
similar picture appears for the auxiliary field at the tricritical
point, $V_{0}^{TCP}$, but its variation is not so abrupt as that
of $h_{0}^{TCP}$, see Fig.~\ref{figc}.

Another important quantity is the magnetization at the TCP. The
equilibrium Eq. (\ref{eqmagn}) at the tricritical point assumes
the form,

\begin{equation}
\frac{\alpha ^{2}F_{3}}{2!}M^{2} + \frac{\alpha ^{5}
F_{6}}{5!}M^{5} = 0 \label{tcpmagn1}
\end{equation}

or

\begin{equation}
F_{6} \omega^{5} + 60 F_{3} \omega^{2} = 0 \label{tcpmagn2}
\end{equation}

\noindent where $\omega \equiv \alpha M$ by taking into account
the conditions for the TCP. The latter equation has the solutions,

\begin{eqnarray}
\omega _{1}^{TCP} & = & 0  \label{tcpmagn0}    \\
\omega _{2}^{TCP} & = & \mbox{\Large(}-60 F_{3} /F_{6}
\mbox{\Large)}^{1/3}       \label{tcpmagn3}
\end{eqnarray}

\begin{figure}[htbp]
\includegraphics*[height=0.20\textheight]{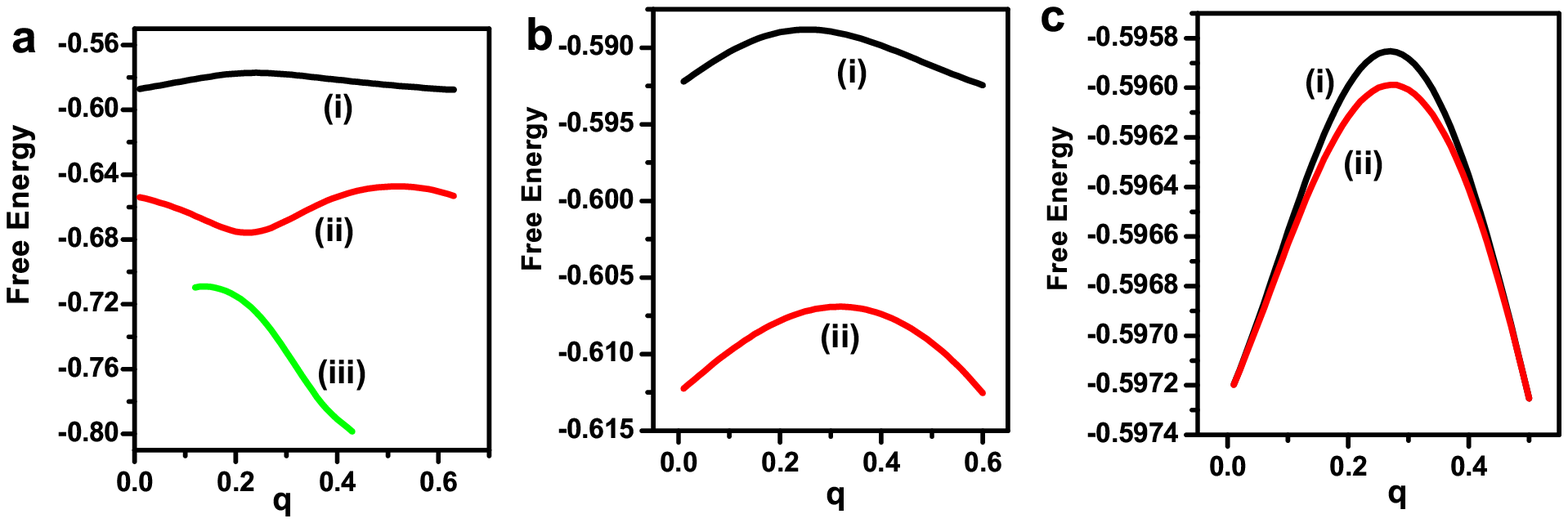}
\caption{\label{figd}(color online) Free energy of the zero
(\ref{tcpmagn0}) and non zero magnetization (\ref{tcpmagn3}) at
the tricritical point for $\lambda = 0.25$ and $p=0.37$ (panel
(a)), $p=0.40$ (panel (b)) and $p=0.50$ (panel (c)). In all
panels, graph (i) corresponds to the zero solution $M_{1}$ and
graphs (ii,iii) to the non zero solutions $M_{2}$ associated with
the upper and lower TCP temperatures, respectively; the zero
magnetization free energy (i) is higher than the one for the non
zero magnetizations (ii,iii), implying that the non zero solution
is the stable one.}
\end{figure}

\begin{figure}[htbp]
\includegraphics*[height=0.25\textheight]{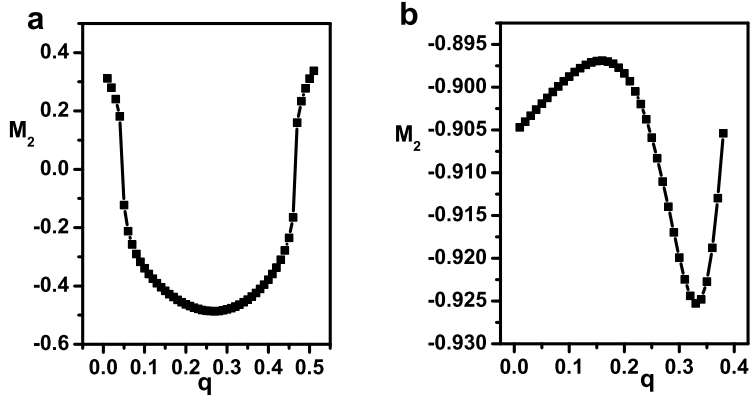}
\caption{\label{fige} The tricritical nonzero
magnetization $M_{2}= \mbox{\large(}-60 F_{3} /F_{6}
\mbox{\large)}^{1/3}$ with respect to q for $\lambda=0.50$,
$p=0.49$ panel (a), and $p=0.62$ panel (b).}
\end{figure}

from which the magnetizations $M_{1,2}^{TCP} = \omega
_{1,2}^{TCP}*(kT^{TCP}/(Jz))$ can be deduced. The non zero TCP
magnetization $M_{2}^{TCP}$ (\ref{tcpmagn3}) has a lower free energy
than the zero solution (\ref{tcpmagn0}), implying that this is the
stable solution at the tricritical point, see Fig.~\ref{figd} for
the respective free energies for $\lambda = 0.25$. In case two
tricritical points exist, then the free energy of the non zero
magnetization corresponding to the lower TCP temperature is
smaller than the respective one for the non zero magnetization
corresponding to the upper TCP temperature for the same $q$
values. The plot of the TCP  magnetization $M_{2}^{TCP}$ appears
in Fig.~\ref{fige}, exhibiting significant variation; however, for
the symmetric probability distribution, $p = q = 0.50$ and
$\lambda = 1$, $M_{2}^{TCP}$ becomes identical with the zero
solution, so that the zero solution, now, is the only one and  
becomes stable for these $p$ and $q$ values; this is, also, 
a result of the elimination of the residual magnetic field implied 
by the first relation in (\ref{h0}). 

The stability of the non zero magnetization over
the zero one is a direct consequence of the existence of the
residual magnetic field due to the first relation in equation
(\ref{h0}), since for the general case $p \neq q$ the mean value
of the random field is non zero, which is equivalent to the
presence of an external magnetic field in the system so that the
magnetization at the tricritical point scales as $M_{t}\equiv
M(T=T^{TCP}) \sim h^{1/\delta_{t}}_{TCP}$, where $h_{TCP}$ is the
random magnetic field and the tricritical exponent $\delta _{t} = 5$ 
according to the Landau theory \cite{asymmetric,asymtrim,stanley,robertson,lawrie}.

In a previous communication \cite{asymmetric}, the PDF of the RFIM
was selected to be the asymmetric bimodal (\ref{bimodalp}); this
system displayed a symmetric behavior at the tricritical point
with respect to the probability $p$\,; especially, two distinct
tricritical points with respective probabilities $p_{1}$ and
$p_{2}$ such that $p_{1}+ p_{2} = 1$ have identical tricritical
temperatures and random fields, whereas the respective auxiliary
fields and non zero magnetizations are absolutely equal. A similar
symmetry is also observed in the present model with respect to the
probabilities $p$ and $q$ for a specific $\lambda$ value; if the
probabilities ($p_{1},q_{1}$) and ($p_{2},q_{2}$) of the modified
trimodal (\ref{trimodalr}) are such that $p_{1} + p_{2}= 1$, then
these two systems have the same TCP temperatures and random
fields, whereas the respective nonzero magnetizations
$M_{2}^{TCP}$ are absolutely equal; the auxiliary fields
$V_{0}^{TCP}$ are absolutely equal in case $\lambda = 1$.
Additionally, these cases have equal the respective free energies
for the zero magnetization ($F(p_{1},q_{1},M_{1}^{TCP}=0) =
F(p_{2},q_{2},M_{1}^{TCP}=0)$) as well as the nonzero ones
($F(p_{1},q_{1},M_{2}^{TCP}) = F(p_{2},q_{2},M_{2}^{TCP})$); the
latter result implies that the two magnetizations
$M_{2}^{TCP}(\lambda,p_{1},q_{1})$,
$M_{2}^{TCP}(\lambda,p_{2},q_{2})$ are equally probable, an
expected result, since the magnetizations have equal absolute
values and the only difference being in their sign so that no
direction is favored  \cite{asymtrim}.

\begin{figure}[htbp]
\begin{center}
\includegraphics*[height=0.50\textheight]{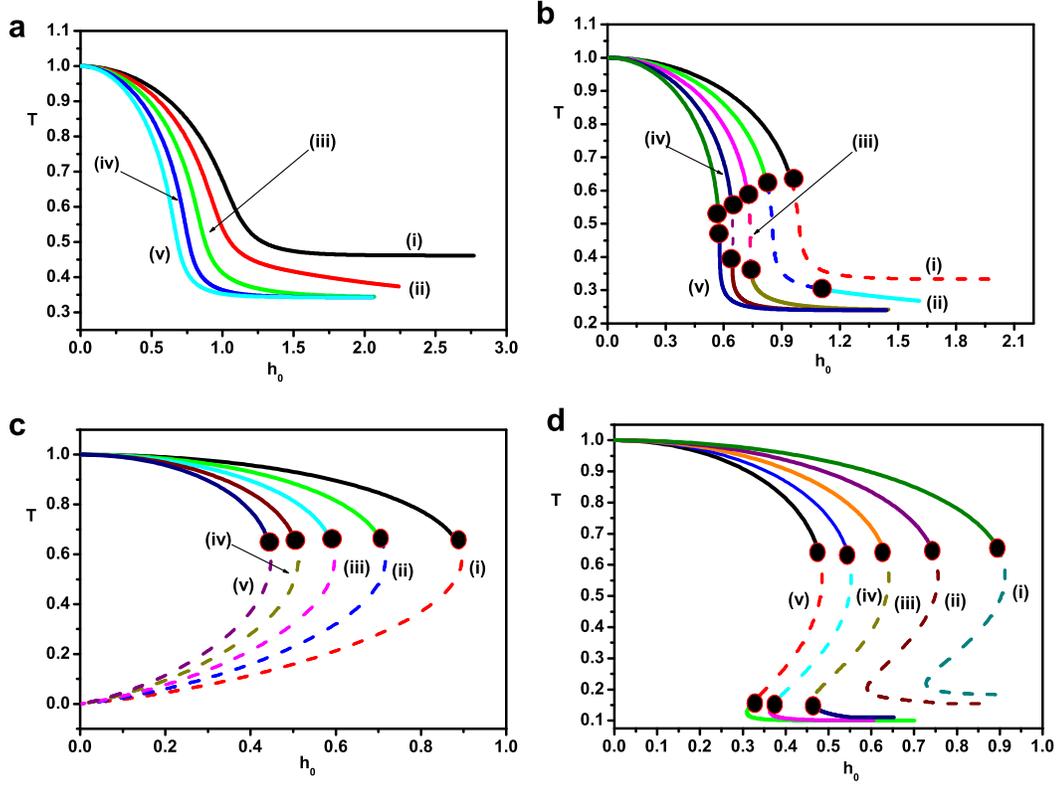}
\caption{\label{figf}(Color online) The phase boundaries for the
PM/FM phase transitions for $\lambda = 0.00(i),0.25(ii),
0.50(iii), 0.75(iv), 1.00(v)$, in all panels; $p=0.35, q=0.30
\, (a)$, $p=0.40, q=0.35 \, (b)$,  $p=q=0.50 \, (c)$, $p=q=0.45 \,
(d)$. A continuous line represents an SOPT and a dashed one an
FOPT joined at a TCP, represented by a full circle. In panels (a,
b) the system is in the FM phase for low temperatures and high
$h_{0}$s; on the contrary, in panel (c) it is in the PM phase for
low temperatures due to re-entrance. In panel (c) the TCP temperatures 
are equal for any value of $\lambda$ ($T^{TCP}=\frac{2}{3}$, see ref. \cite{aharony}), 
whereas in panel (d) the upper TCP temperatures as well as the lower 
ones are not equal among themselves; the difference between the 
respective temperatures of two consecutive graphs is very small. 
Re-entrance is seen in panels (c, d). The temperature $T$ is expressed 
in units of $(Jz/k)$, i.e., $T \equiv kT/(Jz)$.}
\end{center}
\end{figure}

An important component in the study of magnetic or fluid systems
is their phase diagram in which the general behavior of the system
is shown; in the current case, it results as a solution to
Eq. (\ref{magnet1a}) by varying the parameters $p, q, \lambda$ and
appears in the Fig.~\ref{figf} as ($h_{0}-T$) plots labelled by
the individual $\lambda$ value for specific $p$ and $q$ values.
These plots are classified into two main groups: the first one
includes those plots not possessing a TCP and corresponding only
to an SOPT as in the Fig.~\ref{figf}(a) for $p=0.35$, $q=0.30$ for
any $\lambda$ value; the second group includes the plots
possessing at least one TCP, which joins the FOPT branch with the
SOPT branch of the phase diagram as in the Fig.~\ref{figf}(b, c,
d) for $p=0.40$, $q=0.35$, $p=q=0.50$ and $p=q=0.45$,
respectively. In the Fig.~\ref{figf}(b, d) the systems, described
by the corresponding plots, have two TCPs (except the ones for
$\lambda = 0.0$ Fig.~\ref{figf}(b,d) and
$\lambda = 0.25$ Fig.~\ref{figf}(d)); the single and
twin TCPs appear for another values of the parameters $\lambda, p
,q$, as well. In some cases, a random system can present
re-entrance that might be attributed to the competition between
the exchange interaction (ordering factor) due to the first term
in the Hamiltonian (\ref{rham}), on the one hand, and the random
field (disorder factor), on the other hand, as in the
Fig.~\ref{figf}(c, d); this effect is more pronounced in the
Fig.~\ref{figf}(c). In re-entrance a vertical line in the
$(h_{0},T)$-plane crosses the transition line at least twice, in
that, by lowering the temperature at constant $h_{0}$, one
observes firstly a $PM/FM$ transition and then, on further
lowering the temperature, an $FM/PM$ transition appears, so the
magnetization is zero although the temperature is low and the
system remains in the PM phase for low temperatures as in the
Fig.~\ref{figf}(c); occasionally, there can be another PM/FM
transition (additional crossing) with the system returning to the
FM phase for low temperatures and high $h_{0}s$,
Fig.~\ref{figf}(d). Re-entrance appears only in case an FOPT is
present and its direct effect is to limit drastically the extent
of the FM phase space as in the Fig.~\ref{figf}(c). However,
within the MFA re-entrance may lead to nonphysical values
(negative) for the specific heat, since energy will also present
re-entrant behavior as magnetization because energy, within MFA,
is proportional to the magnetization squared thus behaving
similarly. The choice $\lambda=0.00$ in the PDF (\ref{trimodalr})
is a distinct case for this probability distribution, since the
random magnetic field in the ($-z$) direction is eliminated and
it refers to a lattice system in which some of its sites are either
vacant or occupied by non-magnetic particles (site dilution, in this
case the two last terms in (\ref{trimodalr}) become similar and can
be combined to a single term, $P(h_{i})=p\delta (h_{i}-h_{0}) +(q+r)\delta
(h_{i})$), whereas the remaining sites are occupied by magnetic
particles (fraction p) exposed to the random field whose direction
is considered to be the positive z-direction without the presence
of another competing random field; for this case the mean
magnetization is expected to be higher than that for $\lambda \neq
0.00$ for the same $p, q$ as long as there is not competition
between random fields. In all panels of the Fig.~\ref{figf} the
respective phase boundary for $\lambda=0.00$ is the outermost
graph (i) with the widest FM phase region. However, as $\lambda$
is switched on (at constant $p$, $q$) taking on values greater
than zero, the random fields in the negative z direction appear
and oppose the initially prevailing random fields (in the positive
z direction), thus reducing the mean magnetization, causing the
phase boundary to move towards to the temperature axis and,
consequently, reducing the phase space allocated to the FM phase
but simultaneously broadening the PM phase space; the former phase
space attains its smallest extent for $\lambda = 1$, since the
reduction is larger the higher the value of $\lambda$. The plots
in the Fig.~\ref{figf}(a) correspond to systems in the FM phase
for low temperatures and high random fields for any value of
$\lambda$; the critical temperatures seem to tend asymptotically
to the limit $T_{cr} \cong 0.50$ for $\lambda = 0$ and $T_{cr}
\cong 0.35$ for $\lambda \neq 0$, as $h_{0} \rightarrow \infty$.
In the Fig.~\ref{figf}(b) the critical temperatures tend
asymptotically to the limit $T_{cr} \cong 0.35$ for $\lambda = 0$
and $T_{cr} \cong 0.25$ for $\lambda \neq 0$ as $h_{0} \rightarrow
\infty$; in the same limit in the Fig.~\ref{figf}(d) the critical
temperatures tend asymptotically to the limit $T_{cr} \cong
0.1818...$ for $\lambda = 0$ and $T_{cr} \cong 0.1$ for $\lambda
\neq 0$. The succession of phase transitions depends on the number
of the tricritical points present although the SOPTs appear always
for low fields and high temperatures: if there is only one TCP
then the FOPTs appear for high fields and low temperatures;
however, in case two TCPs are present the FOPTs appear for medium
fields and temperatures, whereas the concluding phase transition
is an SOPT for high fields and low temperatures. In the
Fig.~\ref{figf}(b,d) the concluding phase transitions for those
$\lambda$s with two TCPs are of second order and those with one
TCP is of first order as well as in the Fig.~\ref{figf}(c). The
aforementioned reduction of the FM-space due to the gradual
increase of $\lambda$ (control parameter) towards one ($\lambda
\rightarrow 1$) has counterpart in the density profile of a
spherical drop as a function of the inverse range parameter $R$ of
the strength of the attractive forces between the fluid particles;
the overall structure of the density profile remains unchanged as
a function of $R$ in comparison to that with $R=1$ except that the
density profile either shrinks for $R>1$ or widens for $R<1$ to
accommodate inside the drop the available particles
\cite{sphdrop}. A characteristic feature in panel (c), $p=q=0.50$,
is that the TCP temperatures, irrespective of the $\lambda$-value,
are all the same, namely $T^{TCP}=2/3$; this value is identical to
the one estimated by Aharony for the symmetric bimodal PDF
\cite{aharony}.

\begin{figure}[htbp]
\begin{center}
\includegraphics*[height=0.70\textheight]{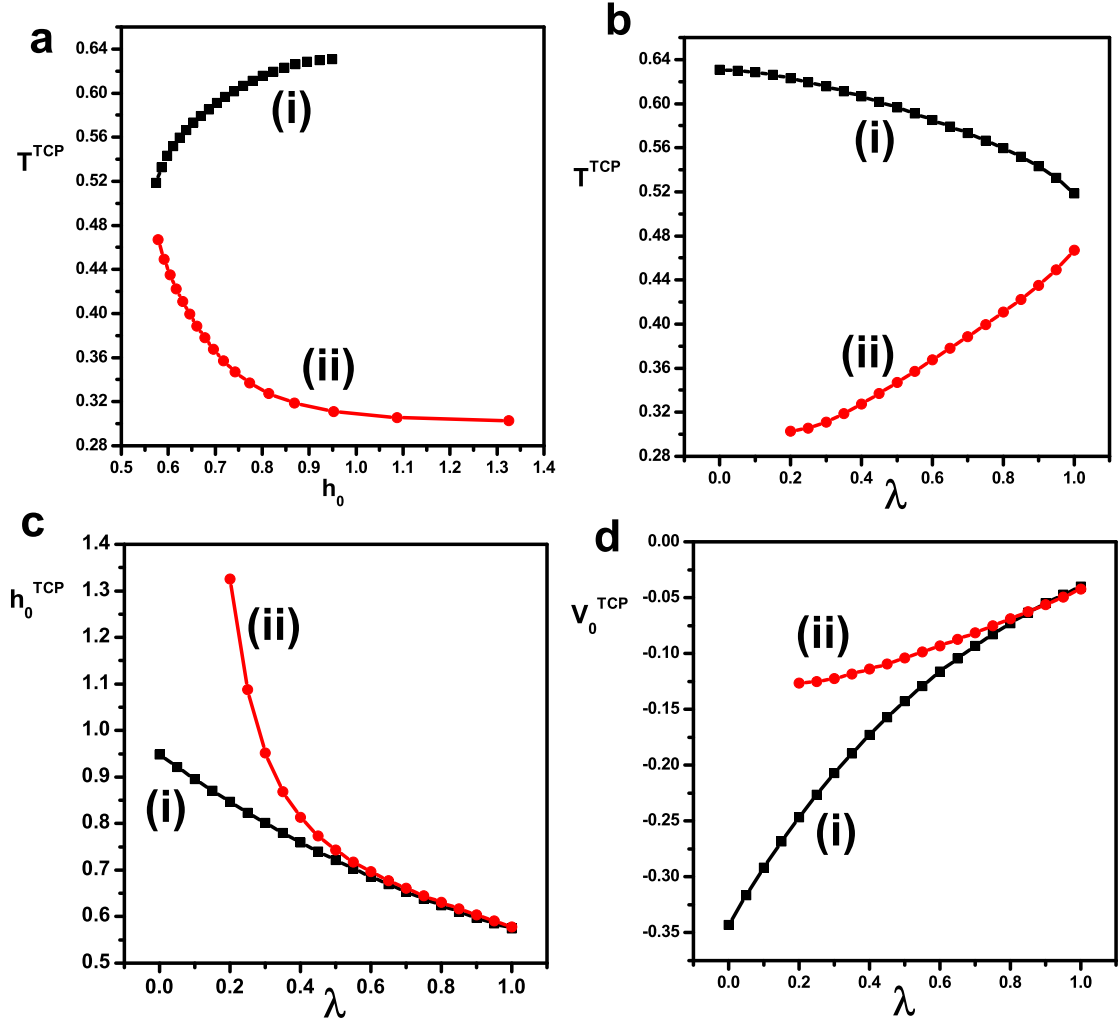}
\caption{\label{figg} (Color online) Tricritical point temperature
$T^{TCP}$ vs $h_{0}$ (panel (a)) as well as TCP coordinates ($T,
h_{0}, V_{0}$) as functions of the competition ratio $\lambda$
(panels (b,c,d)), for $p=0.40, q=0.35$;  the labels (i) (black
symbols) and (ii) (red symbols) refer to the quantities
corresponding to the upper and lower TCP temperatures,
respectively.}
\end{center}
\end{figure}

As it is evident from the Fig.~\ref{figf}(b), both groups of TCP
temperatures regarded as functions of the TCP random field
strength ($h_{0}^{TCP}$) display a systematic behavior, in that,
they follow a decreasing route, which in the last stages (large
$h_{0}$) becomes exponential as is revealed by the graphs in the
Fig.~\ref{figg}(a) resulting by combining both groups of TCP
temperatures. A similar systematic behavior is also followed by
TCP temperature plotted with respect to the competition ratio
$\lambda$, Fig.~\ref{figg}(b); the upper group of temperatures
follow a decreasing route and the lower an increasing route
tending to approach each other as $\lambda$ tends to one, $
\lambda \rightarrow 1$. Similarly, in Fig.~\ref{figg}(c) the
random fields $h_{0}^{TCP}$ as $\lambda$ approaches $1$,
corresponding to the upper group of tricritical temperatures
$(i)$, decrease slowly, whereas those for the lower tricritical
temperatures $(ii)$ initially decrease abruptly but, in the last
stages, decrease slowly tending to approach the ones of the upper
temperatures. In addition, the TCP auxiliary fields $V_{0}^{TCP}$
follow the inverse route in comparison to $h_{0}^{TCP}$: the
$V_{0}^{TCP}$ corresponding to the upper temperatures $(i)$
follows a systematically increasing route, whereas those for the
lower TCP temperatures $(ii)$ are increasing very slowly, tending
to approach the other group of $V_{0}^{TCP}$s as $\lambda
\rightarrow 1$ Fig.~\ref{figg}(d).

An immediate connection of the Fig.~\ref{figa} (describing the
variation of the TCP temperature $T^{TCP}$ with respect to
$p,q,\lambda$) with the Fig.~\ref{figf} (phase diagram) is the
extent of the branches for the SOPTs and FOPTs with respect to these
parameters. According to Fig.~\ref{figa}(b), $T^{TCP}$  initially decreases
implying that the respective SOPT branch increases at the expense
of the FOPT branch acquiring its largest extent when the respective
$T^{TCP}$ has its minimum value for $\lambda=0.25, p=0.38, q=0.21$,
after this point the extent of the SOPT branch starts reducing and
that for the FOPT increasing as the TCP temperature increases;
the reverse behavior of the extent of the branches of SOPTs and FOPTs
occurs for the case of Fig.~\ref{figa}(d) wherein the $T^{TCP}$ initially increases.
In case two such temperatures appear in the phase diagram, as in
Fig.~\ref{figa}(a) and similar plots for other values of the
aforementioned parameters, then the FOPT branch initially
decreases acquiring its smallest extent when $T^{TCP}$ becomes
minimum in the upper branch and maximum in the lower branch, but
later it starts increasing as the two temperatures get farther
apart, whereas according to  Fig.~\ref{figa}(c) in the respective
phase diagram the extent of the FOPT branch continuously increases
as the two TCP temperatures get farther apart with respect to $q$
from the beginning for the right hand part; as far as the left hand
one the SOPT branch, in general, increases.

Solving Eq. (\ref{magnet1a}) to determine the phase diagram, the
magnetization is also calculated for either phase transition. The
condition $A = 1$ or $\alpha F_{2}(\beta,V_{0},h_{0}) = 1$ leads
to

\begin{equation}
  pt_{+}^{2} + qt_{-}^{2} + rt_{0}^{2} = \frac{\alpha -1}{\alpha} \;\;\;\; \;\; \label{t2a}
\end{equation}

and by setting $T_{2} \equiv  p\,t_{+}^{2} + qt_{-}^{2} +
rt_{0}^{2}$, (\ref{t2a}) can be written

\begin{equation}
  T_{2} = \frac{\alpha -1}{\alpha} \;\;\;\; \;\; \label{t2b}
\end{equation}

Inverting Eq. (\ref{t2b}) the respective temperature for either
phase transition can be determined, namely

\begin{equation}
  \frac{kT}{Jz} = 1 - T_{2}   \;\;\;\; \;\; \label{temp}
\end{equation}

In order to specify the type of the transition, the sign of $C
\equiv \alpha^{3} F_{4}/6$ is checked; however, to facilitate the
calculations, the quantity $C$ is rewritten as

\begin{equation}
C = \frac{\alpha^{3}}{3}[4T_{2} - 3T_{4} - 1]
=\alpha^{3}[1-T_{4}-\frac{4}{3\alpha}]  \;\; \;\; \label{consc}
\end{equation}

using (\ref{t2b}) and setting $T_{4} = p\,t^{4}_{+} + q\,t^{4}_{-}
+ rt_{0}^{4}$. For an SOPT, $C$ is negative \cite{aharony}, then
(\ref{consc}) yields

\begin{equation}
T_{4} > 1 - \frac{4}{3\alpha}  \;\;\;\; \;\; \label{sectrns}
\end{equation}

otherwise if

\begin{equation}
T_{4} < 1 - \frac{4}{3\alpha}  \;\;\;\; \;\; \label{firsttrns}
\end{equation}

the resulting transition is an FOPT. In order to determine the
magnetization for an FOPT the expression (\ref{eqmagn}) is
combined with the equality of the respective free energies

\begin{equation}
 F(M=0)  =  F(M\ne 0) \label{feeq1}
\end{equation}

\noindent or,

\begin{equation}
M^{2}=F_{2} \alpha M^{2}+\frac{F_{3}}{3}\alpha^{2}M^{3}+
\frac{F_{4}}{12} \alpha^{3} M^{4}+\frac{F_{6}}{360} \alpha^{5}
M^{6} \label{feeq2}
\end{equation}

\noindent Combining Eqs. (\ref{eqmagn}), (\ref{feeq2}) and using
the condition $\alpha F_{2} =1$, we get

\begin{equation}
F_{6}\omega^{3} + 10F_{4}\omega = 0 \label{fopt2}
\end{equation}

Eq. (\ref{fopt2}), is broken up into two equations; the first is
$\omega _{1} = 0$ or, equivalently, $M_{1} = 0$ for the PM phase,
whereas the other is

\begin{equation}
F_{6}\omega^{2} + 10 F_{4}= 0 \label{fopt3}
\end{equation}

\noindent from which the nonzero solutions result, FM phase,

\begin{equation}
 \omega_{2,3}= \pm \sqrt{-10 F_{4}/F_{6}}   \label{foptsols}
\end{equation}

\noindent with $F_{6} < 0$ as long as $F_{4} > 0$ for an FOPT; the
value for $F_{3}$, consistent with (\ref{eqmagn}) and
(\ref{feeq2}), is $F_{3} = - (F_{4}/6) \sqrt{-10 F_{4}/F_{6}}$ for
the positive root in (\ref{foptsols}) and $F_{3} = (F_{4}/6)
\sqrt{-10 F_{4}/F_{6}}$ for the respective negative root. From the
solution of this equation we can extract the magnetization
$M = \omega*(kT/(Jz))$, since $\alpha=zJ/kT$ is already known from (\ref{temp}).

\begin{figure}[htbp]
\begin{center}
\includegraphics*[height=0.70\textheight]{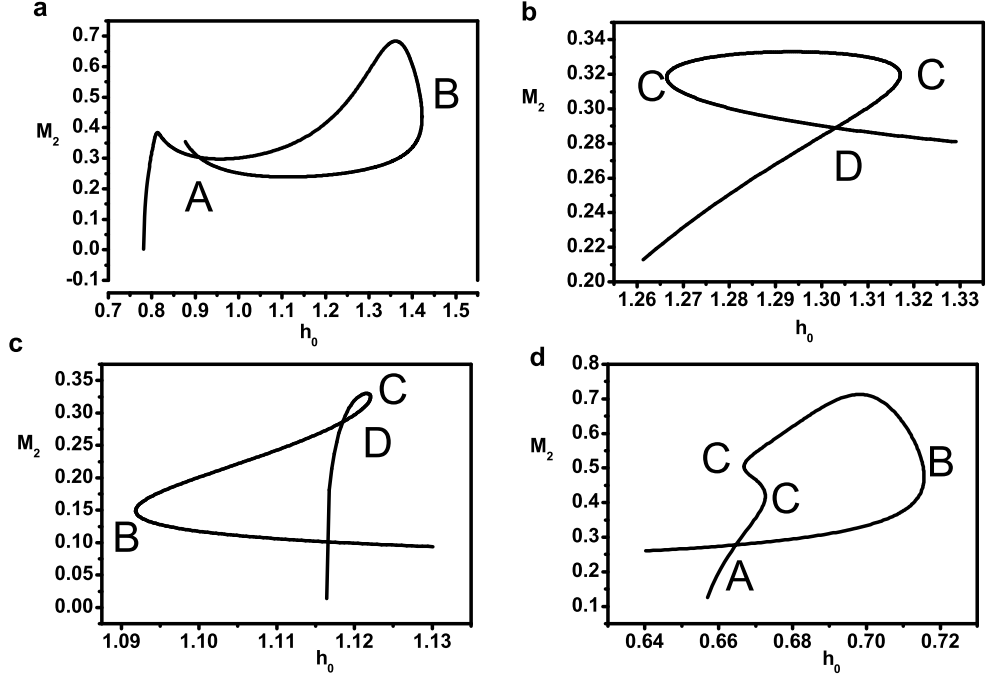}
\caption{\label{figh} Magnetization profile vs. $h_{0}$ for an FOPT.
The first row corresponds to $\lambda = 0.25, p=0.40, q=0.50$ panel (a) and
 $\lambda = 0.50$, $p=0.25, q=0.55$ panel (b). The second
row corresponds to $\lambda = 0.50, p=0.30, q=0.45$ panel (c)
and $\lambda = 0.50, p=0.40, q=0.50$ panel (d). In
all panels point A represents a double critical point, point B a
regular critical point, point C a critical end-point, point D a
double critical end-point.}
\end{center}
\end{figure}

In plotting the phase diagram or the order parameter profile, the
temperature is usually chosen as the independent variable;
however, this is not the only choice as any other variable,
suitably chosen, can be. In the present case the randomness
strength $h_{0}$ is considered to be the control parameter for
studying the variation of the non-zero positive magnetization
$M_{2} = \omega_{2}*(kT/(Jz))$ for an FOPT, Eq. (\ref{foptsols}),
by forming the respective magnetization profile as a function of
$h_{0}$ for specific values of $\lambda$, $p$ and $q$; the
negative solution $M_{3} = - M_{2}$ behaves analogously. This
study reveals a complicated structure for magnetization profiles
and we present some representative of them; they appear in
Fig.~\ref{figh} displaying a variety of structures and
characterized by critical points of several kinds. Apart from the
simple profiles, with or without a critical point,
there are also profiles forming closed loops (one or two), closed
miscibility gap, Fig.~\ref{figh}. In addition to the
usual critical points (indicated by the letter B in all graphs),
there are double critical points (A points), critical end-points
(C points) as well as double critical end-points (D points)
\cite{asymtrim,hadjievans}. This behavior can be considered as an
FOPT between the two coexisting $M_{2}$ magnetizations with
respect to randomness $h_{0}$ as long as this is now the control
parameter, with upper and lower critical temperatures.

We consider, now, the values of $p$ and $q$ for which the system
exhibits only an SOPT; Eq. (\ref{magnet1a}) takes the form for
$A=1$,

\begin{equation}
F_{6}\omega^{5} + 20F_{4}\omega^{3} + 60F_{3}\omega^{2} = 0
\label{sopt1}
\end{equation}

The value $\omega_{1} = 0$ is again a solution (two-fold) or,
equivalently, $M_{1} = 0$ (PM phase); the other three ones are the
solutions to the equation

\begin{equation}
F_{6}\omega^{3} + 20F_{4}\omega + 60F_{3} = 0 \label{sopt2}
\end{equation}

\noindent which, depending on the value of $\lambda, p, q$ and
$h_{0}$, can have either only one real non zero solution if
$\Delta = u^{3} + v^{2} \geq 0$ ($v = -30F_{3}/F_{6}$,
$u=(20F_{4})/(3F_{6})$), namely

\begin{equation}
\omega_{2}=\sqrt[3]{v+\sqrt{\Delta}} + \sqrt[3]{v-\sqrt{\Delta}}
\label{root2}
\end{equation}

\noindent or three real non zero solutions for $\Delta < 0$, which are

\begin{eqnarray}
\omega_{2} & =& 2 \sqrt[3]{\rho}\;\cos(\theta/3)    \nonumber   \\
\omega_{3} & = & -\sqrt[3]{\rho}\;[\cos(\theta/3) +
\sqrt{3}\sin(\theta/3)]                            \nonumber \\
\omega_{4} & = & -\sqrt[3]{\rho}\;[\cos(\theta/3) -
\sqrt{3}\sin(\theta/3)]                            \label{root34}
\end{eqnarray}

\noindent where $\rho = \sqrt{v^{2} - \Delta}$, $\theta = \arctan(
\sqrt{-\Delta}/v )$ and $M_{i} = \omega_{i}*(kT/(Jz))$, $i=2,3,4$.
As a consequence, the solutions for an SOPT are classified into
two groups, group 1 includes the zero-solution ($M_{1}=0$) and the
single nonzero one $M_{2}$ of Eq. (\ref{root2}), whereas group 2
includes again the zero solution and the nonzero ones $M_{2},
M_{3}, M_{4}$ of the Eq. (\ref{root34}). Depending on the values
of $\lambda, p, q$ and $h_{0}$, there can be transitions between
these two groups. For the $p$ values an FOPT is present, the
solutions to the SOPT Eq. (\ref{sopt2}) belong to group 1; for
small $h_{0}$'s the zero solution ($M_{1}=0$) is the stable,
whereas for larger $h_{0}$'s (but smaller than those corresponding
to the respective FOPT) the $M_{2}$ solution (\ref{root2}) is the
stable.

The investigation was also extended to the zero-temperature case,
$T=0$; in this case the free energy (\ref{mfafren}) reduces to,

\begin{eqnarray}
F \equiv \frac{1}{N}\langle F \rangle_{h} & = & \frac{1}{2}
zJM^{2} - \frac{1}{\beta} \langle \ln\{ 2 \cosh [\beta (z J M +
h_{i})] \}
\rangle_{h}          \nonumber  \\
  &  =  &  \frac{1}{2} zJM^{2} - \langle |z J M + h_{i}| \rangle_{h}    \nonumber  \\
  & = &  \frac{1}{2} zJM^{2} - p\, |z J M + h_{0}| -q\, |z J M - \lambda*h_{0}|-r\, |zJM|  \label{ztfren}
\end{eqnarray}

\noindent the external potential was omitted. Applying the
equilibrium condition $dF/dM = 0$ to (\ref{ztfren}) we get,

\begin{equation}
M = p\, \frac{| zJM + h_{0} |}{zJM + h_{0}} + q\, \frac{| zJM -
\lambda*h_{0} |}{zJM - \lambda*h_{0}} + r\, \frac{| zJM|}{zJM}
\label{ztmagnet}
\end{equation}

Analyzing Eq. (\ref{ztmagnet}), we find that $M = 1$ is a stable
solution for $p+r > \lambda h_{0}/zJ$, whereas for $p+r <
 \lambda h_{0}/zJ$ the stable one is $M = 1 - 2q$. Also, across the
boundary $ \lambda h_{0}/zJ = p+r$ a first-order phase transition
occurs between the two ordered phases with $M = 1$ and $M = 1 -
2q$.  If we consider the symmetric trimodal PDF
($p=q=\frac{1-r}{2}, \lambda =1 $), the results found by
Sebastianes and Saxena \cite{saxena} are recovered, that is, the
former result ($M = 1$) is stable for $ \frac{1+r}{2}> h_{0}/zJ$,
whereas the latter ($M = r$) for $\frac{1+r}{2} < h_{0}/zJ$, using
the current notation. The physical explanation for the existence
of the above two ordered phases ia that they can be attributed to
the competition between the ordering tendency, due to the first
term in Eq. (\ref{rham}), and the disorder induced because of the
presence of the second term in the same equation. In the first
case, $M = 1$, the condition $(p + r) > (\lambda h_{0}/(zJ))$
implies that the exchange interaction $J$ is much stronger than
the randomness $h_{0}$, and their ratio is always smaller than
one, thus forcing the system's spins to order according to the
first term in (\ref{rham}). The alternative condition $(p + r) <
(\lambda h_{0}/(zJ))$ implies, now, that randomness is no longer
negligible but strong enough to influence significantly the spins
enforcing a $p$-fraction of them to point up and a $q$-fraction
down, to randomly align with the local fields, thus, practically,
it dominates, so to speak, over the first term in Eq. (\ref{rham})
so that $M = p-q+r = 1 - 2q$. In addition to the aforementioned
two solutions, there are more; the result $M=2(p+q)-1$ is stable
for $p-r>\lambda h_{0}/zJ$, $M=2p-1$ for $ p-r < \lambda
h_{0}/zJ$, $M=1-2p$ for $ r-p > \lambda h_{0}/zJ$, $M=1-2(p+q)$
for $ r-p < \lambda h_{0}/zJ$ and $M=-1$ for $p+r+\lambda
h_{0}/zJ>0$.

\vspace{-6mm}

\section{Conclusions and discussions}

\vspace{-6mm}

In the current treatment we have determined the phase diagram and
discussed some critical phenomena of the Ising model under the
influence of a trimodal random field, an extension of the bimodal
one to allow for the existence of non magnetic particles or
vacancies in the system, for arbitrary values of the probabilities
$p$ and $q$ and different strengths of the random field in the up
and down directions specified by the competition parameter
$\lambda$ via the Landau expansion. The competition between the
ordering effects and the randomness induces a rich phase diagram.
The system is strongly influenced by the random field, which
establishes a new competition favoring disorder; this is obvious
from the appearance of first order transitions and tricritical
points, in addition to the second order transitions for some
values of $\lambda, p$ and $q$; the tricritical point temperature
has various modes of variation as a function of $p$ and $q$ and
for some cases there are two such points. The trimodal
distribution induces re-entrant behavior for the appropriate range
of $p, q$ and random field $h_{0}$. For some values of $p$ and $q$
the system can be found either in the PM phase or in the FM phase
for low and medium temperatures and high random fields; a
significant result is that the part of the phase diagram allocated
to the FM phase is reduced significantly as $\lambda$ tends to
one. A direct consequence of the asymmetric and anisotropic PDF is
the existence of residual mean magnetic field in the system, a
result of $<h_{i}>_{h} = (p-\lambda \,q)h_{0}$, making the TCP non
zero magnetization $M_{2}$ to be the stable one in comparison to
the zero one, $M_{1}$. Both asymmetric PDFs, bimodal and trimodal,
confirm the existence of a TCP and, nevertheless, yield similar
magnetization profiles as well as re-entrance; however, the
trimodal one predicts also the existence of a second TCP.

Griffiths extended the notion of the critical point to the
so-called multicritical points, e.g., the tricritical point,
the critical-end-point, double critical-end-point, fourth-order point,
ordered critical point, etc. \cite{griffiths}; however, in order
to describe these points (except the first two) the expansion considered
for the free energy (\ref{mfafren2}) has to be extended
to higher-order terms \cite{trimodal,crok3,galambirman,milmanetal}
so that the stability criteria for such a point are satisfied, but
this is beyond the scope of the current research.

The Landau theory breaks down close to the critical point (the non
classical region) because as the transition temperature is
approached the fluctuations become important and non classical
behavior is observed. A relative criterion, called the Ginzburg
criterion, determines how closely to the transition temperature
the true critical behavior is revealed, or, in other words, it
governs the validity of the Landau theory close to a critical
point \cite{ginzburg}. This criterion can rely on any
thermodynamic quantity but the specific heat is usually considered
for determining the critical region around $T_{c}$ where the mean
field solution cannot correctly describe the phase transition. The
Landau theory is valid for lattice dimensionality greater than the
upper critical dimension $d_{u}=4$ in the case of the presence of
only thermal fluctuations. However, in the current case the
presence of random fields enhances fluctuations causing the
critical region to be wider than the one due only to the thermal
fluctuations \cite{kaufmankardar,nielsen} and the upper critical
dimension is increased by $2$ to $d_{u}=6$. Occasionally, the non
classical region for some physical systems is extremely narrow so
that the respective critical behavior expected from Landau theory
is observed for a wide range of temperatures because, in this
case, the fluctuation region is very narrow and hardly accessible
for experimental observation; such a system is the weak-coupling
superconductor in three dimensions for which the respective non
classical region is $|t_{CR}|\leq 10^{-16}$ ($t_{CR}$ is the
reduced temperature, $t_{CR}=(T-T_{c})/T_{c}$). However, on
reducing the space dimension as in the case of the weak-coupling
superconductor in two dimensions, the non classical region
expands, so that the critical exponents have their classical
values up to the interval $|t_{CR}|= 10^{-5}$, and thus the
reduction of the space dimensionality has serious consequences for
the critical behavior of the physical system; in contrast, there
are systems with a wide non classical interval as in the case of
the superfluid helium transition, for which the classical region
extends up to $|t_{CR}|= 1.0$ and so fluctuations are detectable
\cite{patapok,ivan,domb,goldenfeld}. In addition to
superconductivity, the extent of the non classical region for the
ferroelectric system triglycine sulfate (TGS) is relatively small
and its critical exponents have the respective classical values up
to $|t_{CR}|= 1.5\times 10^{-5}$ \cite{fe1,fe2,fe3}.

Our results indicate that on increasing the complexity of the
model system new phenomena can be revealed as in the current case
of including asymmetry in the PDF; this inclusion induces drastic
changes in the phase diagram, such as re-entrance and two TCPs,
thus confirming the necessity of treating the partial
probabilities $(p,q,r)$ of the PDF in the most general way to get
the complete phase diagram. A similar situation appears in the
model systems in Refs. \cite{galamaharony,galam1} wherein the
complexity considered has revealed a rich variety of phase
diagrams with known and new multicritical points. The results
obtained in the current investigation by using the MFA can provide
a basis for a comprehensive analysis as well as experimental
implementation. However, they are of no less importance, since
they nevertheless show the phenomena that expected to be observed.

\vspace{-9mm}

\ack{This research was supported by the Special Account for
Research Grants of the University of Athens ($E\Lambda KE$) under
Grant No. 70/4/4096.}


\newpage

\begin{thebibliography}{00}

\bibitem{rsdim}I. A. Hadjiagapiou, A. Malakis, S. S. Martinos, Physica A 387 (2008) 2256.

\bibitem{rbim} I. A. Hadjiagapiou I A, Physica A 390 (2011) 1279.

\bibitem{huiberker} K. Hui, A. Nihat Berker, Phys. Rev. Lett. 62 (1989) 2507.

\bibitem{physicstoday} D. S. Fisher, G. M. Grinstein, A. Khurana,
Phys. Today 41 (12) (1988) 56.

\bibitem{natermannvillain} T. Nattermann, J. Villain, Phase Transit. 11 (1988) 5.

\bibitem{imryma} Y. Imry, S.-K. Ma, Phys. Rev. Lett. 35 (1975) 1399.

\bibitem{blume} M. Blume, Phys. Rev. 141 (1966) 517.

\bibitem{capel}  H. W. Capel, Physica 32 (1966) 966.

\bibitem{maletal} A. Malakis, A. Nihat Berker, I. A. Hadjiagapiou, N. G. Fytas,
 Phys. Rev. E 79 (2009) 011125.

\bibitem{malakisetal} A. Malakis, A. Nihat Berker, I. A. Hadjiagapiou, N. G. Fytas,
T. Papakonstantinou, Phys. Rev. E 81 (2010) 041113.

\bibitem{imbrie} Imbrie J Z, Phys. Rev. Lett. 53 (1984) 1747.

\bibitem{aharony} A. Aharony, Phys. Rev. B 18 (1978) 3318.

\bibitem{sneiderpytte} T. Schneider, E. Pytte, Phys. Rev. B 15 (1977) 1519.

\bibitem{fernandez} L. A. Fern\'{a}ndez, A. Gordillo-Guerrero, V. Mart\'{\i}n-Mayor,
 J. J. Ruiz-Lorenzo, Phys. Rev. Lett. 100 (2008) 057201.

\bibitem{gofman} M. Gofman, J. Adler, A. Aharony, A. B. Harris, M. Schwartz,
Phys. Rev. Lett. 71 (1993) 2841; Phys. Rev. B 53 (1996) 6362.

\bibitem{houghton} A. Houghton, A. Khurana, F. J. Seco, Phys. Rev. B 34 (1986) 1700.

\bibitem{galambirman2} S. Galam, J. L. Birman, Phys. Rev. B 28 (1983) 5322.

\bibitem{machta} J. Machta, M. E. J. Newman, L. B. Chayes, Phys. Rev. E 62 (2000) 8782.

\bibitem{middleton} A. A. Middleton, D. S. Fisher, Phys. Rev. B 65 (2002) 134411.

\bibitem{fytas1} N. G. Fytas, A. Malakis, K. Eftaxias, J. Stat. Mech. (2008) P03015.

\bibitem{hernandezetal} L. Hern$\acute{a}$ndez, H. T. Diep, Phys. Rev. B 55 (1997) 14080;
L. Hern$\acute{a}$ndez, H. Ceva, Physica A 387 (2008) 2793.

\bibitem{fishaha} S. Fishman, A. Aharony, J. Phys. C: Solid State Phys. 12 (1979) L729.

\bibitem{galam3} S. Galam, Phys. Rev. B 31 (1985) 7274.

\bibitem{kaufkan} M. Kaufman, M. Kanner, Phys. Rev. B 42 (1990) 2378.

\bibitem{andelman1} D. Andelman, Phys. Rev. B 27 (1983) 3079.
\bibitem{dgaussian} N. Crokidakis, F. D. Nobre, J. Phys.: Condens. Matter 20 (2008) 145211.

\bibitem{galamaharony} S. Galam, A. Aharony, J. Phys. C: Solid St. Phys. 13 (1980) 1065.

\bibitem{galam1} S. Galam, J. Phys. C: Solid St. Phys. 15 (1982) 529.

\bibitem{galam2} S. Galam, Europhys. Lett. 37 (1997) 615; J. of Non-Crystalline Solids 235-237 (1998) 570.

\bibitem{asymmetric} I. A. Hadjiagapiou, Physica A 389 (2010) 3945.

\bibitem{asym2} I. A. Hadjiagapiou, Physica A 390 (2011) 3204.

\bibitem{trimodal} M. Kaufman, P. Klunzinger, A. Khurana, Phys. Rev. B 34 (1986) 4766.

\bibitem{saxena} V. K. Saxena, Phys. Rev. B 35 (1987) 2055; R. M. Sebastianes, V. K. Saxena, Phys. Rev. B 35 (1987) 2058.

\bibitem{asymtrim} I. A. Hadjiagapiou,  Physica A 390 (2011) 2229.

\bibitem{riegeryoung} H. Rieger, A. Peter Young, J. Phys. A: Math. Gen. 26 (1993) 5279.

\bibitem{rieger} H. Rieger, Phys. Rev. B 52 (1995) 6659.

\bibitem{hartmannyoung} A. K. Hartmann, A. P. Young, Phys. Rev. B 64 (2001) 214419.

\bibitem{nowak} U. Nowak, K. D. Usadel, J. Esser, Physica A 250 (1998) 1.
\bibitem{dukovski} I. Dukovski, J. Machta, Phys. Rev. B 67 (2003) 014413.

\bibitem{malakisfytas} A. Malakis, N. G. Fytas, Phys. Rev. E 73 (2006) 016109;
 Eur. Phys. J. B 50 (2006) 39.

\bibitem{belangeryoung} D. P. Belanger, A. P. Young, J. Magn. Magn. Mater. 100 (1991) 272.

\bibitem{khurana} A. Khurana, F. J. Seco, A. Houghton, Phys. Rev. Lett. 54 (1985) 357.

\bibitem{crok2} N. Crokidakis, F. D. Nobre, Phys. Rev. E 77 (2008) 041124.

\bibitem{crok3} O. R. Salmon, N. Crokidakis, F. D. Nobre, J. Phys.: Condens. Matter 21 (2009) 056005.

\bibitem{galam} S. Galam, C. S. O. Yokoi, S. R. Salinas, Phys. Rev. B 57 (1998) 8370.

\bibitem{weizenmann} A. Weizenmann, M. Godoy, A. S. de Arruda,
D. F. de Albuquerque,N. O. Moreno, Physica B 98 (2007) 297.

\bibitem{stanley} H. Eugene Stanley,  Introduction to Phase Transitions and Critical
Phenomena,  Oxford, Clarendon Press, 1971.

\bibitem{robertson} H. S. Robertson, Statistical Thermophysics, New Jersey,
Prentice-Hall, 1993, pp. 303, 308.

\bibitem{lawrie} I. D. Lawrie, S. Sarbach, Theory of Tricritical Points,
in Phase Transitions and Critical Phenomena, ed. C. Domb and J. L.
Lebowitz, Academic Press, London, U.K., Vol. 9, 1984.

\bibitem{sphdrop} I. Hadjiagapiou, J. Phys.: Condens. Matter 7 (1995) 547.

\bibitem{hadjievans} I. Hadjiagapiou, R. Evans, Mol. Phys. 54 (1985) 383.

\bibitem{griffiths} R. B. Griffiths, Phys. Rev. B 12 (1975) 345.

\bibitem{galambirman} S. Galam, J. L. Birman, J. Phys. C : Solid State Phys. 16 (1983) L1145.

\bibitem{milmanetal} F. S. Milman, P. R. Hauser, W. Figueiredo, Phys. Rev. B 43 (1991) 13641.

\bibitem{ginzburg} V. L. Ginzburg, Fiz. Tverd. Tela. (Leningrad) 2
(1960) 2031 [Sov. Phys.-Solid State 2 (1961) 1824].

\bibitem{kaufmankardar} M. Kaufman, M. Kardar, Phys. Rev. B 31 (1985) 2913.

\bibitem{nielsen} J. Als-Nielsen, R. J. Birgeneau, Amer. J. Phys. 45 (1977) 554.

\bibitem{patapok} A. Z. Patashinskii, V. I. Pokrovskii, Fluctuation Theory of
Phase Transitions, Pergamon Press, Oxford, U.K., 1979.

\bibitem{ivan} Y. M. Ivanchenko, A. A. Lisyansky, Physics of Critical Fluctuations,
 Springer-Verlag, New York, U.S.A, 1995.

\bibitem{domb} Domb C, The Critical Point: A Historical Introduction to the
Modern Theory of Critical Phenomena, Taylor and Francis, London, U.K., 1996.

\bibitem{goldenfeld} Goldenfeld N, Lectures On Phase Transitions And The Renormalization Group,  Addison-Wesley, Reading, U.S.A., 1992.

\bibitem{fe1} K. Deguchi, E. Nakamura, Phys. Rev. B 5 (1972) 1072.

\bibitem{fe2} A. Mercado, J. A. Gonzalo, Phys. Rev. B 7 (1973) 3074.

\bibitem{fe3} T. Mitsui, E. Nakamura, M. Tokunaga, Ferroelectrics 5 (1973) 185;
M. Tokunaga, T. Mitsui, Ferroelectrics 11 (1976) 451.


\end{thebibliography}
\end{document}